\documentstyle[11pt,epsfig]{article}
\def\baselinestretch{1.3}
\newcommand{\ba}{\begin{array}}
\newcommand{\ea}{\end{array}}
\newcommand{\bd}{\begin{displaymath}}
\newcommand{\ed}{\end{displaymath}}
\newcommand{\be}{\begin{equation}}
\newcommand{\ee}{\end{equation}}
\newcommand{\bea}{\begin{eqnarray}}
\newcommand{\eea}{\end{eqnarray}}

\def\q2 {q^2}

\begin{document}
\begin{flushright}
hep-ph/0407225
\end{flushright}

\begin{center}
{\Large\bf Sparticle spectrum and phenomenology in split supersymmetry:
some possibilities}\\[20mm]
Biswarup Mukhopadhyaya\footnote{E-mail: biswarup@mri.ernet.in}\\
{\em Harish-Chandra Research Institute,\\
Chhatnag Road, Jhusi, Allahabad - 211 019, India}\\
Soumitra SenGupta \footnote{E-mail: tpssg@iacs.res.in} \\
{\em Department of Theoretical Physics,\\
Indian Association for the Cultivation of Science\\
Jadavpur, Kolkata- 700 032, India}
\\[20mm]
\end{center}
\begin{abstract}
We investigate the split supersymmetry (SUSY) scenario recently proposed by
Arkani-Hamed and Dimopoulos, where the scalars are heavy but
the fermions are within the TeV range.  We show that the sparticle spectrum
in such a case crucially depends on the specific details of the mechanism
underlying the SUSY breaking scheme, and  the accelerator signals 
are also affected by it. In particular, we demonstrate in the context of a
braneworld-inspired model, used as illustration in the original work, 
that a new fermion $\psi_X$, arising from the SUSY breaking
sector, is shown to control low-energy phenomenology in several cases.
Also, SUSY signals are characterised
by the associated production of the light neutral Higgs.
In an alternative scenario where the gauginos
are assumed to propagate in the bulk, we find that gluinos can be
heavy and short-lived, and the SUSY breaking scale can be free of 
cosmological constraints.
\end{abstract}

\vskip 1 true cm
\newpage
\setcounter{footnote}{0}
\def\baselinestretch{1.8}

Most attempts at phenomenological modeling of a supersymmetric nature
envision supersymmetry (SUSY) breaking at the energy scale of a TeV or
below in the observable sector of elementary particles. This is motivated
by an aspired solution to the  naturalness problem. However, it has
been recently argued \cite{arkani} that SUSY may have nothing to
do with the naturalness
problem, although it may be the necessary ingredient of a fundamental 
theory of nature
like string theory where SUSY is most often a necessary ingredient.
Such an argument has its origin in the observation that a broken SUSY
introduces a cosmological constant proportional to the fourth power of the
SUSY breaking scale, thereby requiring fine-tuning of a much more severe degree
than what is required to make the Higgs boson mass free of  quadratic
divergence. Appeal has therefore been made to the
so-called `landscape'  scenario \cite{susskind},
where a large multitude of universes exist, corresponding to different vacua of
string theory. It is thus statistically feasible \cite{douglas} for us,
it is argued, to live
in a universe fine-tuned in the way we have it, thereby having both a
small cosmological constant and an electroweak scale stabilised in the
100 GeV range.

A recently proposed SUSY scenario \cite{arkani}, inspired by such a
philosophy, has postulated  that the
SUSY breaking scale may be as large as $10^{10} -10^{13}$ GeV, having
all scalar masses (excepting that of the light neutral Higgs) of such
high values. However, it can have fermions (gauginos/Higgsinos) with masses
within the TeV range. While the heavy scalars provide a natural
suppression of flavour-changing neutral currents, the light fermions
can (a) present an appropriate cold dark matter candidate in the form
of the lightest SUSY particle (LSP), and (b) help in the unification
of the three forces \cite{giudice}. Although it has been
suggested \cite{calmet} that this is not absolutely necessary for
Grand Unification, such a `split SUSY' scenario is nonetheless
of considerable interest, both from theoretical and phenomenological
viewpoints \cite{models}. In this note we wish to point out a few
phenomenological possibilities which may have implications
in accelerator experiments.

We demonstrate these possibilities with reference to a particular SUSY
breaking scheme discussed in reference [1], based on a braneworld scenario.
This scenario is stabilised by a new chiral supermultiplet (X) whose fermionic
component ($\psi_X$) turns out to be as light as the gauginos and Higgsinos.
We illustrate the different situations that arise corresponding to the
different relative masses of these fermions, resulting in  different
types of SUSY phenomenology. Furthermore, we suggest that some variants
of such a model, where the gauginos are kept in the bulk rather than confined
to the visible brane, can make the gauginos heavier and cause them (for
example, the gluino) to decay quite fast. This alleviates the
cosmological problems associated with a  long-lived gluino while
the field $\psi_X$ can still be a viable dark mater candidate.

	The model consists of five-dimensional supergravity (SUGRA)
compactified  on $S_{1}/Z_2$ orbifold with radius r \cite{sugra}.
Apart from the SUGRA multiplet the
low energy effective action at the tree level includes the radion chiral
superfield $T = r + \theta^2 F_T$ and the chiral compensator field
$\phi = 1 + \theta^2 F_{\phi}$ \cite{chiral}.
It is further assumed that there is a constant
superpotential $c$ on one of the two branes (the one {\em not} containing
the standard model superfields)
which are located at the two orbifold fixed points \cite{constant}. 
This scenario
is shown to break supersymmetry in a way which is equivalent to
the coordinate dependent compactification scheme proposed by 
Scherk and Schwarz \cite{scherk}.
Such constant superpotential which may originate from gluino
condensation\cite{horava} in the hidden sector on the brane, gives rise to a non-vanishing
vacuum expectation value (vev) of the auxiliary component of $T$ leading
to the breakdown of supersymmetry with the gravitino mass
given by $m_{3/2} = c/r$.
Although the cosmological constant vanishes at the tree level, the one-loop
gravitational Casimir energy gives rise to a negative potential which goes to
a minimum for $r=0$. To make the braneworld stable, $N$ bulk hypermultiplet of
mass $M$ are
included which because of their repulsive Casimir energy stabilises the
minima of the potential at $r = M^{-1}$. However, the minimum of the potential
being still negative, one introduces a chiral superfield $X$ with
non-vanishing $F_X$ to make the cosmological constant zero.
More precisely, one requires

\begin{equation}
F(X)~\simeq~m^2~\simeq~{\frac{1}{4\pi r^2}}
\end{equation}

This chiral superfield X is localized on the same brane where the standard
model superfields also reside.
Together with the  radion superfield T and the chiral compensator
field $\phi$ (both propagating in the bulk), X gives
contributions to the Kahler potential and the superpotential, which are
given by

\begin{equation}
K = \phi^{\dagger}\phi \left[ X^{\dagger}X +
{\frac{a_1}{M^2_5}}(X^{\dagger}X)^2 +
higher~powers\right]
\end{equation}

\noindent
and
\begin{equation}
 W =  \phi^3 m^2 X
\end{equation}

\noindent
where the five-dimensional Planck mass is related to the four-dimensional
Planck mass by

\begin{equation}
M^2_4~=~\pi r M^3_5
\end{equation}

Using the above model, different terms in the SUSY Lagrangian generate
masses for the sfermions, gauginos, Higgsinos as well as the scalar and
spinor components of X. Although these are indicated in reference [1], we
summarise in table 1 the exact expressions for the different mass terms
that arise from the corresponding terms in the Lagrangian. It may be
noticed that we have used some general coefficients in our Kahler potential
as well as in the interaction terms. The purpose of doing so is to
convey the message that the strengths of these terms are
{\it prima facie} unrelated, and that the different masses induced by them
can be ordered in different ways in a general scenario. These
different parameters can appear, for example, in a string-inspired scenario
via the vev's of the moduli of the compact dimensions.

The `split' character of the (s)particle spectrum here has to do with
the fact that\\ (a) $\langle X \rangle~\sim~m^2/M_5$ while
$\langle F_X \rangle~\sim~m^2$, and
(b) $m~\sim~1/r$ while $r$ relates the four-and five-dimensional Planck
masses as in equation (4). One needs to have $M_5$ in the range
$10^{16-17}$ GeV in order to have a SUSY dark matter candidate within
the TeV scale \cite{arkani}. Also, the universal gaugino mass parameter
derived above is
subject to the usual running before the different gaugino masses at the
electroweak scale are obtained.

\begin{center}
$$
\begin{tabular}{|c|c|c|c|c|}
\hline
 &Origin in ${\cal L}_{SUSY}$  & Expression\\
 & &  \\
\hline
 $m_{\tilde{f}}$ & $\int d^4 \theta \alpha_0 X^{\dagger}X Q^{\dagger}Q/M^2_5$ &
$\alpha_0 \pi M^5_5/M^4_4$\\
  & & \\
\hline
  $m_{X}$ & $ \int d^4 \theta a_1 (X^{\dagger}X)^2/M^2_5$  &
$a_1^{1/2} \pi M^5_5/2\sqrt{2}M^4_4$ \\
  & & \\
\hline
  $m_{\psi_X}$ & $\int d^4 \theta a_1 (X^{\dagger}X)^2/M^2_5$ &
$a_1 \pi^2 M^9_5/12M^8_4$     \\
  & & \\
\hline
  $M_{1/2}$ & $\int d^2 \theta \alpha_1 m^2 X W_{\beta} W^{\beta}/M^3_5 + h.c.$     &     \\
   & + $\int d^4 \theta \alpha_2 X^{\dagger} X W_{\beta} W^{\beta}/M^3_5$    &
$(\alpha_1 + \alpha_2) \pi^2 M^9_5/M^8_4$    \\
  & & \\
\hline
  $\mu$ & $\int d^2 \theta \alpha_3 m^2 X H_u H_d/M^2_5$    &     \\
        &+ $ \int d^4 \theta \alpha_4 X^{\dagger}X H_u H_d/M^2_5$   &     \\
        &+  $ \int d^4 \theta \alpha_5 m^2 X^{\dagger} H_u H_d/M^3_5$ & $(\alpha_3 + \alpha_4
+ \alpha_5) \pi^2 M^9_5/M^8_4$   \\
  & & \\
\hline
\end{tabular}
$$
\end{center}
{\em Table 1: The different terms in the SUSY Lagrangian
contributing to the sfermion and fermion masses including the
Higgsino mass parameter $\mu$. The corresponding expressions for the
masses derived are also shown, in terms of the different coefficients
and the power-law dependence on the four-and five-dimensional Planck scales.}\\

\noindent
where W can be expressed in terms of the components of a gauge supermultiplet
as
\begin{equation}
W_{\alpha}~=~4i\lambda_{\alpha} - \left[4 \delta^{\beta}_{\alpha} D
+ 2i (\sigma^{\mu} \bar{\sigma}^{\nu})^{\beta}_{\alpha} V_{\mu\nu}\right]
\theta_{\beta} + 4 \theta^2 \sigma^{\mu}_{{\alpha}{\dot{\alpha}}}
\partial_{\mu}\bar{\lambda}^{\dot{\alpha}}
\end{equation}

\noindent
D and $V_{\mu\nu}$ being respectively the auxiliary part of
the superfield and the field strength of the gauge boson
$V_\mu$.

Our next step is to examine the different possible SUSY spectra in the
fermionic sector, and outline the way they are likely to affect SUSY cascades
at collider experiments. Before we enter into such an analysis, it is important
to mention a few points.

First, this scenario admits of  Higgsino-Higgs-$\psi_X$ couplings of the form

\begin{equation}
{\cal L}~=~(\alpha_3~+~\alpha_4) {\frac{\pi}{4}} ({\frac{M_5}{M_4}})^4 \bar{\psi}
\tilde{H_u}H_d~+~h.c.
\end{equation}

When electroweak symmetry is broken, such interaction also leads
to Higgsino-$\psi_X$ mixing which, however, is suppressed by
$(\frac{M_5}{M_4})^4$. It can be easily verified that such small
off-diagonal terms in the neutralino mass matrix are of little
consequence in determining the composition of states, and therefore
the states obtained in the corresponding MSSM case remain valid here as well.

Secondly, the above interaction term can lead to the decay
$\psi_X \longrightarrow \tilde{H} h^0$ where  $h^0$
is the lightest neutral scalar, for $m_{\psi_X} > (\mu~+~M_{h^0})$.
Alternatively, a heavier Higgsino can decay into the Higgs and $\psi_X$.
The decay width for any of these modes is driven by
the suppressed coupling given above. However, one finds that
for a Higgs mass of 120 GeV, and with the heavier(lighter) of the
two fermions being of mass 250 (100) GeV, the lifetime
varies between $10^-2$ and $10^{-10}$ seconds for
${\frac{M_5}{M_4}}$ varying between  $10^{-3}$ and $10^{-2}$. Such
lifetimes are small compared to the cosmological scale, but
can sometimes lead to displaced vertices in detectors.

Thirdly, the coupling of a gaugino to a
gauge boson and $\psi_X$ arises from the
terms proportional to $\alpha_1$ and $\alpha_2$ in table 1, and is of the form

\begin{equation}
{\cal L}~=~2(\alpha_1 ~+~ \alpha_2)\pi {\frac{M_5^3}{M_4^4}}
\bar{\psi}\sigma^{\mu\nu}\lambda V_{\mu\nu}~+~h.c.
\end{equation}

This interaction is heavily suppressed; the fact that it arises from
a dimension-five term involving $V_{\mu\nu}$ never allows the decay
to recover from  suppression by the Planck mass.
 Therefore, the gluino has to decay via squark propagators, and is
long-lived when the latter involve large masses. The electroweak gauge bosons,
however, have unsuppressed gauge couplings with the Higgs superfields and
therefore allow quick decays for them if kinematically allowed.

Let us now consider the different ways the masses of the electroweak gauge
boson, the Higgsinos and $\psi_X$ can be ordered. Various possibilities can
arise here depending on the values of the parameters
$a_1$ and $\alpha_i$ (i = 1-5). Based on the observations listed above,
these possibilities can lead to different types of SUSY phenomenology
discussed below.
We are assuming that the gluino is on the heavier side; in any case it
will remain long-lived in all cases since it does not couple to
Higgsinos.

\begin{enumerate}
\item $\underline{M_{gaugino} > M_{\psi_X}> \mu.}$ Here the gluino is
long-lived as expected, but not the electwroweak gauginos, which have
unsuppressed decays into $h^0$ and a Higgsino. Therefore, the electwroweak
gauginos (which are within experimentally accessible range), once produced in
colliders, can  decay quickly into the Higgsino LSP which is also
the dark matter candidate. The field $\psi_X$, too, is within the TeV range,
and can also decay into the LSP fast enough, but its production in
experiments is suppressed, as shown in  equations (6) and (7). The
conclusion, therefore, is that  $\psi_X$ does not play any significant role
in superparticle cascades. Such cascades end up in a Higgsino LSP,
accompanied by the production of the lighter neutral scalar $h^0$ if
it can be produced on-shell, or  a $b\bar{b}$ pair otherwise
\footnote {In the case of the charged gaugino $\tilde{W}^+$, the cascade also
associates the production of real or virtual W's.}.

\item $\underline{M_{gaugino}> \mu  > M_{\psi_X}.}$ Here, together with a
long-lived gluino, the electroweak gauginos have the unsuppressed decay
into the Higgs-Higgsino pair, where the latter, however, is liable to
go again to an $h^0$ (real or virtual) and the $\psi_X$. The time scale for
the second decay is in the range $10^-2$ - $10^{-10}$ s, as mentioned earlier.
{\em Thus  $\psi_X$ marks the culmination of all SUSY cascades, and is
actually the LSP as well as the dark matter candidate, thus
governing the final-state kinematics
whenever superparticles (except gluinos) are produced in colliders.
This is the case where  $\psi_X$ enters seriously into accelerator
phenomenology}.
Again, the simultaneous appearance of one or more light neutral Higgses
is a characteristic feature of SUSY signals. For charged gauge
bosons/Higgsinos, the final states also include either the W-boson or
pairs of fermions typical of charged Higgs decays.

\item $\underline{ M_{\psi_X} > M_{gaugino}> \mu.}$ Obviously, $\psi_X$
has extremely suppressed production rate and has no
role in phenomenology here. The electroweak gauginos (unlike the gluino)
again decay immediately into the $h^0$ and the Higgsino LSP.

\item $\underline{ M_{\psi_X} > \mu > M_{gaugino}.}$ This is a situation
similar to the previous one, excepting that it has a gaugino-dominated
LSP this time. The Higgsinos appear via direct production or
in cascades but decay fast into the $h^0$ and LSP.
The heavier electroweak gauginos (namely, the lighter chargino $\chi^{\pm}_1$
and the second lightest neutralino $\chi^0_2$) in this case also decay
into the LSP, but such decays are controlled by the small overlap between
the LSP and the neutral SU(2) gaugino $\tilde{W}_3$. Again, a sizeable
fraction of SUSY cascades in collider experiments includes real or 
virtual $h^0$ in the final state.

\item $\underline{\mu > M_{gaugino} > M_{\psi_X}.}$ Here the Higgsino states
have unsuppressed decays into the electroweak gauginos (the gluino as usual is
long-lived). The neutral Higgsino, of course, can still have the decay
into the $\psi_X$, but such a decay has a very small branching ratio compared
to that into gauginos, as seen from equation (6). The decay of the
lightest neutralino, on the other hand, is suppressed by the Planck mass.
Thus the $\psi_X$-state, although it is the LSP, is never reached within a measurable
time in SUSY cascades, and is decoupled from phenomenology. This situation is in
contrast with that in case 2 above, where the  $\psi_X$ has a role to play as
the LSP. The lightest neutralino, as a quasi-stable particle, actually
turns out to be a parallel dark matter candidate; also the other charginos and
neutralinos decay into it, thereby making it the effective LSP.

\item $\underline{\mu  > M_{\psi_X}  > M_{gaugino}.}$ Again, the phenomenology
is controlled by the Higgsinos and gauginos, with the $\psi_X$ having
practically no role in experimental signals. All fermions excepting the
gluino and the $\psi_X$ are short-lived. The LSP is gaugino-dominated and
is a dark matter candidate, but so also is  $\psi_X$, being quasi-stable due
to equation (7).
\end{enumerate}

Finally, we consider a special case, where the gauginos propagate in the bulk and
explore whether this helps to reduce the lifetime of the gluino.
It has been shown in reference [1] that a long-lived gluino
imposes a constraint on the SUSY breaking scale. This is because the
presence of extremely long-lived gluinos could lead to the
formation of abnormally heavy isotopes. Therefore, one may demand that
the gluino lifetime be smaller than the age of the nucleosynthesis era. 
As gluino decays involve heavy
squark propagators, an upper limit on the squark masses and consequently
on the SUSY breaking scale can therefore be obtained.  
In principle, equation (7) also leaves room for the decay
$\tilde{g} \longrightarrow g \psi_X$.
A quick look at equation (7), however, tells us that this decay is suppressed
by the Planck scale in the denominator.
We show that this suppression can be offset if the gauginos
exist in the bulk, and consequently acquire large masses.

When gauginos propagate in the bulk, they can couple to the radion field as \cite{luty}

\begin{equation}
L = \int d^2\theta T W_{\alpha}W^{\alpha} +h.c
\end{equation}
Such interaction yields a gaugino mass of the form
\begin{equation}
M_{gaugino}~ =~ <F_T>/M_5~ =~m^2/M_5
\end{equation}

If now the rate for such a gluino decaying into a gluon and the $\psi_X$
is calculated from the
gluino-$\psi_x$-gauge boson coupling as given in equation (7),
we find after using equations (1) and (4) that
the lifetime of a gluino is on the order of  $10^{-10}$ seconds!
Thus the gluino has a fast enough decay to avoid all cosmological
problems, irrespective of what the squark masses are. The
constraint derived in reference [1] on the SUSY breaking scale $M_S$,
namely, $M_{S} \le 10^{13}$ GeV for a TeV-scale gluino, is no 
longer applicable, and SUSY
can be broken at even higher scales. Of course, a scenario of this kind
will decouple all the gauginos from accelerator experiments. The only
low-energy signatures of SUSY  will come from the Higgsinos (and
the field $\psi_X$ if its mass happens to be lighter than $\mu$). 
It is, however, obvious that the gauginos in such a case become far too 
heavy to provide the right threshold for gauge coupling unification.
Thus a bulk gaugino scenario of the above kind helps to avoid the
problem of long-lived hadrons at the cost of sacrificing the gauge coupling
unification via supersymmetry. To achieve grand unification in this scenario
some other new physics will have to exist
\cite{calmet}.

In conclusion, we have explored the consequences of a split supersymmetry
scenario, and shown that details of the mechanism underlying the
splitting between the scalar and fermion masses can have a crucial
role to play in low-energy SUSY phenomenology. We have demonstrated this
with special reference to a SUGRA-based scenario used by Arkani-Hamed
and Dimopoulos, where a chiral superfield X plays a pivotal role. We have
examined the different types of mass spectra that can emerge in the process.
In particular, the spinor component of X is the LSP in some of the scenarios,
and its role in the collider signatures of SUSY can be noticeable.
Another remarkable prediction is that {\em in practically all the cases
the observed SUSY signals will be accompanied by the production of the
lightest neutral Higgs boson}.
Finally, we have shown that an alternative scenario where the gauginos
are assumed to propagate in the bulk leads to heavy  but short-lived gluinos,
whereby the upper limit on the SUSY breaking scale disappears.

{\bf Acknowledgment:} BM acknowledges the hospitality of Indian Association
for the Cultivation of Science, Kolkata, where this study was initiated. SSG
thanks Harish-Chandra Research Institute for hospitality during the
concluding part of the project.


\begin{thebibliography}{99}
\bibitem{arkani} N.Arkani-Hamed and S.Dimopoulos, hep-th/0405159.
\bibitem{susskind} S.Weinberg, Phys. Rev. Lett. {\bf 59}, 2607(1987);
R.Bousso and J.Polchinski,  JHEP {\bf 0006}, 006 (2000);
L.Susskind, hep-th/0302219; M. Douglas, JHEP 0305,046 (2003);
S.Kachru, R.Kallosh, A.Linde and S.P. Trivedi, Phys. Rev. D 68, 046005(2003).
\bibitem{douglas} M.R. Douglas, hep-th/0405279.
\bibitem{giudice} G.F. Giudice, A. Romanino, hep-ph/0406088.
\bibitem{calmet} X. Calmet, hep-ph/0406314.
\bibitem{models} A.Arvanitaki, C.Davis, P.W. Graham and
J.G. Wacker, hep-ph/0406034 ; A.Pierce, hep-ph/0406144;
S.Profumo, C.E. Yaguna, hep-ph/040703; S.Zhu hep-ph/0407072;
M.Dine, E.Gorbatov , S.Thomas, hep-th/0407043;
P. H. Chankowski, A. Falkowski, S. Pokorski, and J. Wagner, hep-ph/0407242;
M.Cvetic, P.Langacker, T.Li, T. Liu., hep-th/0407178;
E.Silverstein, hep-th/0407202.
\bibitem{sugra}  D.Marti and A.Pomarol,  Phys.Rev.{\bf D64} 105025 (2001);
D.E.Kaplan and N.Weiner, hep-th/0108001;; M.Luty and N.Okada,  JHEP {\bf 0304}, 050 (2003).
\bibitem{chiral} W.D.Linch, M.Luty and J.Phillips, Phys. Rev. {\bf D68}
025008 (2003)
\bibitem{constant} J.Bagger, F.Feruglio and F.Zwirner, JHEP {\bf 0202}, 010
(2002).
\bibitem{scherk} J.Scherk and J.H. Schwarz, Nucl.Phys. {\bf B153}, 61 (1979).
\bibitem{horava}  P.Horava, Phys. Rev. D54,7561(1996).
\bibitem{luty} Z.Chako and M.Luty, hep-ph/0008103; T.Kobayashi and K.Yoshioka,
  Phys. Rev. Lett {\bf 85}, 5527 (2000).
\end{thebibliography}
\end{document}